\documentclass[a4paper]{jpconf}
\usepackage{graphicx}
\usepackage[square,sort&compress]{natbib}
\usepackage{amsmath,amssymb}

\begin{document}
\title{Signatures of Dark Matter in Cosmic-Ray Observations}

\author{Alessandro Cuoco}

\address{Institute for Theoretical Particle Physics and Cosmology, RWTH Aachen University, Sommerfeldstr.~16, 52056 Aachen, Germany}

\ead{cuoco@physik.rwth-aachen.de}

\begin{abstract}
I provide  a short review of the current status of indirect dark matter searches with gamma rays, charged cosmic rays and neutrinos.
For each case I  will focus on various excesses reported in the literature  which have been interpreted as possible hints of dark matter, and I will use them as examples to discuss theoretical aspects and analysis methodologies.
\end{abstract}

\section{Introduction}

We have strong  evidence for the existence of dark matter (DM) at galactic and cosmological scales.
Nonetheless, all the evidences probe the gravitational effects of DM, while a probe of the particle nature
of DM is so-far still missing. Searching for the particle nature of DM typically involves three different approaches (see e.g.~\cite{Plehn:2017fdg}),
i.e., searches at colliders, direct detection, and indirect detection.
In the latter, subject of this short review, 
one can look at DM signatures in the astrophysical environments where DM is abundant and 
reveal them through visible Standard Model (SM) messengers like electromagnetic radiation, cosmic rays (CR) or neutrinos.
A popular DM model is given by weakly interacting massive particles (WIMPs), which are well motivated theoretically since they 
naturally explain the observed DM abundance  (the so-called WIMP miracle). Furthermore, they also offer striking signatures in all the 3 channels.
However, it's easy to build models which have strong signatures in one of the 3 channels but none in the other two (see e.g.,~\cite{Feng:2010gw}).
Another well-studied candidate is given by   axions, which offer multiple ways for direct detection,
as well as different indirect signatures, but cannot be seen in colliders.

For indirect detection, the main assumption is that the DM particle can annihilate or decay
into standard model particles either directly or through a mediator, so that the secondary SM final particles
can be detected and reveal the presence of DM.
In the last years, due to new precise observations, several `anomalies'
in the various indirect channels have appeared which could be interpreted as a signature of DM.
In the following, I will focus on these hints and use them as example to illustrate the
main underlying methodologies and concepts.

\section{Gamma Rays}\label{sec:gamma}

DM searches in gamma rays have been revolutionized in the last 10 years by the advent of the Fermi
Large Area Telecope (LAT), which provided a large statistic of data in the energy range $\sim$ 100 MeV - 1 TeV
with very precise angular and energy resolution. 
The region above TeV is instead covered by ground bases observatories like HESS, VERITAS, MAGIC, HAWC
and the forthcoming CTA observatory.
This allowed a large number of different studies
based on the analysis of different targets, which I will discuss below.

\subsection{Milky Way Dwarf galaxies}\label{sec:dwarfs}

Dwarf satellite galaxies orbiting the Milky Way (MW) are DM-dominated objects with
a very sub-dominant baryonic content, given mainly by stars. Thus, they are ideal objects
to look for gamma rays from DM with a very low astrophysical background expected.
The intensity of the DM gamma-ray signal in the case of DM annihilation is characterized by the so-called J-factor:
\begin{equation}\label{eq:jfactor}
J_{\Delta \Omega} =  \int\limits_{\Delta \Omega}\!\textrm{d}\Omega \! \int\limits_{\textrm{l.o.s}}\!\textrm{d}s \rho^2{\left(r(s,\theta)\right)}\,.
\end{equation}
i.e., the integral along the line of sight of the square of the DM density profile $\rho(r)$. 
This assumes a typical  vanilla model in which DM annihilates in s-wave so that
the quantity $\langle \sigma v \rangle$ (i.e., the thermally averaged DM annihilation cross-section) is velocity independent and proportional
to the above J-factor.
At present, observations of dwarf satellite galaxies of the MW provide the most stringent limits
on $\langle \sigma v \rangle$ \cite{Ackermann:2015zua,Fermi-LAT:2016uux}. The constranits are shown in Fig.~\ref{fig:CCEBenito}.  
Models in which DM annihilates not in s-wave are of course possible, and in this case the constraints on 
$\langle \sigma v \rangle$ will depend on the velocity distribution of DM and not only on the DM distribution profile.

The main  uncertainty comes from   the not perfectly know DM distribution and overall DM content in the dwarfs,
which has to be inferred using the observed stars as kinematic tracers of the DM profile.
This gives an uncertainty in the J-factor which is taken into account in the 
constraints of  \cite{Ackermann:2015zua,Fermi-LAT:2016uux}.
Nonetheless, some simplifying assumption is typically used, as, e.g, isotropy of the velocity field of the stars. 
Adopting more conservative assumptions, e.g. \cite{Bonnivard:2015xpq}, the limits can be weakened by a factor up to a few.
Constraints from dwarfs will improve in the next years  especially thanks to new dwarfs which are expected to be discovered
in future galaxy surveys~\cite{Charles:2016pgz}.

\subsection{The galactic center excess}\label{sec:gce}

\begin{figure}[t]
\vspace{-1cm}
\centering
\includegraphics[width=0.58\textwidth]{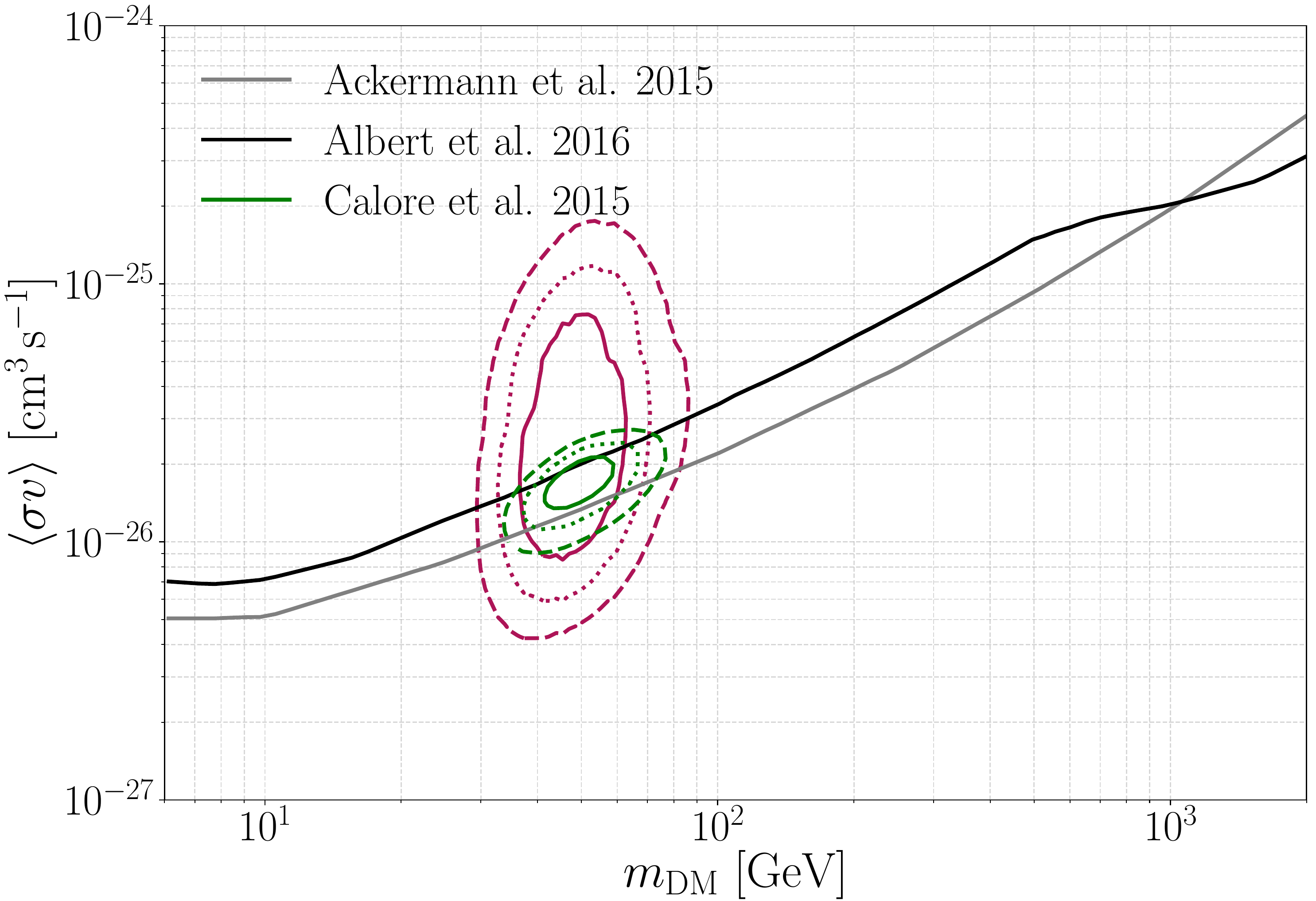}
\caption{1, 2 and 3 $\sigma$ contours in the  $(m_{\rm DM}, \langle\sigma v\rangle)$ plane for the $b\bar{b}$ 
DM annihilation channel from the analysis  of \cite{Benito:2019ngh} (large contours), and
from the work of \cite{Calore:2014xka} (smaller contours).    Also included are the upper limits from the analysis of
Milky Way dwarf galaxies \cite{Ackermann:2015zua,Fermi-LAT:2016uux}.
}
\label{fig:CCEBenito}
\end{figure}

The presence of a Galactic Center Excess (GCE) has been reported by several groups in 
the last few years~\cite{Abazajian:2012pn,Gordon:2013vta,Daylan:2014rsa,Calore:2014xka,TheFermi-LAT:2015kwa}. 
By `excess' is meant an excess over the MW diffuse gamma-ray emission, which is produced
by CRs interacting with gas and the radiation field in the Galaxy.
The excess is at the level of 10-20\% of the MW diffuse.
Nonetheless, the conclusion of the above studies is that it cannot be accommodated in the standard
MW diffuse even taking into account the related uncertainties, and it thus require a different explanation.
The GCE seems compatible with a spherical morphology,  extending up to~10$^\circ$ away from the galactic center,   
with a steep `cuspy' radial profile~\cite{Daylan:2014rsa,Calore:2014xka}, and with
an energy spectrum  peaked at a few GeV.
Various astrophysical mechanisms and scenarios have been proposed to explain the excess~\cite{Petrovic:2014uda,Petrovic:2014xra,Cholis:2015dea}. 
On the other hand, intriguingly, it has been shown that the 
excess is also compatible with an interpretation
in terms of DM annihilation, with a cross-section close to the thermal  value and with a DM mass around 50~GeV.
Fig.~\ref{fig:CCEBenito} shows the region preferred by the DM interpretation of the GCE using as GCE spectrum and morphology the one
derived in~\cite{Calore:2014xka}. The two different set of contours will be discussed in the next section.

In 2016 a fatal blow to the DM interpretation of the excess seemed to come from two separate studies
of the clumpyness of the GCE \cite{Bartels:2015aea,Lee:2015fea}.  In the case of DM origin the GCE should be smooth, while if due to astrophysical
contributions, in particular to a population of unseen unresolved astrophysical sources, the excess should be
quite clumpy. The two studies  \cite{Bartels:2015aea,Lee:2015fea} designed an analysis to quantify the
clumpyness and both found that the excess is too clumpy to be due to DM, thus favoring the astrophysical interpretation
of the excess. 
However, a recent study \cite{Leane:2019xiy} scrutinized in detail the analysis of \cite{Lee:2015fea} and found that it was not robust,
mainly because of a not accurate enough description of the underlying MW diffuse emission, i.e., the 
background for the GCE determination. The DM interpretation of the excess seems thus, at the moment,
viable again.  It's unclear if also the study \cite{Bartels:2015aea} suffers of the same problem
of \cite{Lee:2015fea}. A systematic study like the one of \cite{Leane:2019xiy}, but relative to \cite{Bartels:2015aea} would be desirable to clarify the issue.

On a longer term, a robust way to establish if the GCE is due DM is via  dwarfs observations.
In the next years, dwarfs constraints should became strong enough to robustly test the excess~\cite{Charles:2016pgz}.
If the GCE in DM is real, then a similar signal in dwarfs will have to appear.
A possibility to disprove the GCE DM interpretation would be to detect the population
of unresolved sources, in the radio band\cite{Calore:2015bsx} or in gravitation waves~\cite{Calore:2018sbp}.
Finally, a new space-based gamma-ray detector with a significantly improved angular resolution
w.r.t. Fermi-LAT would also be able to clearly identify the sources, if present, and to strongly reduce 
the systematic uncertainties in separating the excess from the diffuse background.

\subsection{Dark matter density profile and uncertainties}\label{sec:Jfactor}

All the various DM probes searching for DM signatures in the MW rely on the knowledge
of the MW DM profile and its normalization.
On the other hand, the DM density in the MW is  measured only in the vicinity of the solar system, and only with a quite large
uncertainty, mostly systematic in nature. In the inner galaxy, in particular, no direct measurements are available since the gravitational
potential is dominated by the baryonic matter.  Extrapolations are thus necessary together with assumptions about the
shape of the DM density profile.
The uncertainties in the MW DM profile thus play an important role in the
interpretation of DM signatures. 
This  uncertainty  and its impact on indirect DM searches and, in particular, on the GCE
has been studied in  \cite{Benito:2019ngh}.
In \cite{Benito:2019ngh},  a generalized Navarro-Frenk-White (NFW) profile 
\begin{equation}\label{eq:nfw}
\rho(r) =   \rho_s    \left(\frac{r}{r_s}\right)^{-\gamma}   \left(1+\frac{r}{r_s}\right)^{-3+\gamma},
\end{equation}
has been employed, which is generic enough to catch the uncertainties related to DM distribution.
The parameters of the gNFW profile have been constrained using the MW rotation curve data, i.e., kinematical
tracers of the DM potential in the MW.
The derived uncertainties in the DM profile, quantified through  the parameters, $\gamma$, $\rho_s$ $r_s$ and $R_0$ ($R_0$ being the distance of the
Solar system from the GC, which indirectly also enters in the determination of the profile)
as well as further uncertainties in the distribution of baryonic mass in the Galaxy, have been taken into account to provide GCE DM contours marginalized over all
these uncertainties. The result are \mbox{shown in Fig.\ref{fig:CCEBenito}}. It can be seen that the contours of the GCE taking into 
account the uncertainties in the DM profile are much bigger than the ones derived in  \cite{Calore:2014xka} assuming a fixed profile,
showing the importance of the role played by these uncertainties.
Also, it's shown that the --anyway minor-- tension between the GCE and constraints from dwarfs galaxies  is completely removed
when the uncertainty in the MW DM distribution is accounted for.

\subsection{Other gamma-ray probes}
Dwarfs galaxies and the Galactic Center are not the only targets which can be searched for a DM gamma-ray signal.
Other probes include galaxy clusters,    unidentified gamma-ray sources (which could be  dark MW satellites or not known yet dwarfs),
the MW halo, the extra-galactic gamma-ray sky, gamma-ray lines and axion signatures. 
For a review and discussion on future prospect see \cite{Conrad:2017pms,Charles:2016pgz}. 
An interesting probe, in particular, is the extra-galactic gamma-ray sky background (EGB), since it offers 
several ways to search for a DM signal. It differs from the other probes since in this case the observable
is an entire map as function of energy, rather than one or more targets in given directions on the sky.
Thus,  the traditional EGB DM searches, which make use of the EGB energy spectrum only,
can be complemented by techniques which analyze the statistical properties of the map,
as the auto-correlation function, the cross-correlation with gravitational tracers of DM, like galaxy catalogs
or lensing shear surveys, and the 1-point probability distribution function statistic.
A review of the subject is given in \cite{Fornasa:2015qua}. See also \cite{Fornasa:2016ohl,Cuoco:2015rfa}.

\subsection{DM in x-rays }

Moving away from the gamma-ray band, DM can leave signatures in the electromagnetic spectrum
at basically all wavelengths. In the radio band, DM can leave its imprint through electrons and positrons
produced through DM annihilation/decay, which then produce a radio signal via synchrotron emission
in the astrophysical magnetic fields. See, for example \cite{Colafrancesco:2005ji,Fornengo:2011xk,Buch:2015iya}.
X-ray radiation can be produced by the same above $e^+e^-$ but this time via inverse Compton on the 
ambient background low energy (radio-microwave-infrared) radiation.
An interesting alternative mechanism, not involving WIMPs, is given by sterile neutrinos of $\sim$ KeV mass.
They represent a viable DM candidate and can two-body decay into an active neutrino and a photon
thus giving a line in the x-ray band.
Indeed, an anomalous line in the spectrum of galaxy clusters has been identified at about 3.5 KeV, which could be interpreted
as originating from a sterile neutrino of $\sim$ 7 KeV \cite{Bulbul:2014sua,Boyarsky:2014jta}.
The DM interpretation of the line should be clarified by new  upcoming high resolution x-ray
observatories. See \cite{Boyarsky:2018tvu} for a detailed discussion.

\subsection{CMB signatures}

Finally, a further indirect probe of DM, not strictly categorizable in the above discussed classification, 
is its signature on the cosmic microwave background (CMB) and
in particular on its anisotropies. In this case, DM annihilation or decay during the recombination epoch
can alter the ionization history of the universe and, in turn, affect CMB anisotropies. 
Constraints derived from this effect can be quite strong especially for low ($\lesssim 10$ GeV) DM 
masses~\cite{Slatyer:2015jla,Aghanim:2018eyx,Poulin:2016anj,Galli:2009zc}.

\section{Cosmic Rays}

Since DM is neutral under SM charges,  it's expected to produce via decay/annihilation  equal amounts of particles and
antiparticles.  Since antimatter is rare in our universe, 
searches for DM in charged CRs are performed looking at spectra of antimatter particles, like positrons and
antiprotons, where the astrophysical background is low.
An obvious drawback with respect to gamma rays is that charged CRs do no propagate in straight lines from the source
and thus propagation in the Galactic magnetic field needs to be modeled and included to predict the DM signal.
Nonetheless, the signal to background ratio is quite favorable, and this makes this channel quite promising.
A recent review is given in~\cite{Gaggero:2018zbd}.

\subsection{Positrons}

A rising in energy of the positron ($e^+$) fraction above $\sim 10$ GeV was first clearly measured
by PAMELA~\cite{Adriani:2008zr} and later confirmed by AMS-02~\cite{Aguilar:2013qda} with high precision.
The fact that a rising $e^+$ fraction is not expected in the standard Galactic CR model, 
where positrons are secondaries produced in collisions of CR nuclei with ambient gas,  
caused a  great deal of excitement  that a primary positron component, in particular from DM, could be present. 
The DM interpretation, however, requires $\sim$ TeV DM mass and very high cross-sections
$\sim 10^{-23}$cm$^{-2}$s$^{-1}$, and it's thus in contrast with several complementary observations,
in particular, dwarfs galaxies~\cite{Lopez:2015uma}, the MW halo~\cite{Ackermann:2012rg} and CMB constraints~\cite{Slatyer:2015jla,Aghanim:2018eyx}.
A DM origin of CR $e^+$ is thus at the moment very disfavored.
A likely explanation is given by pulsars, which are indeed expected to produce positrons,
and fit well  the current $e^+$ and $e^+ + e^-$spectrum~\cite{DiMauro:2017jpu}.

\subsection{Antiprotons}

CR antiprotons have been widely investigated as a tool to search for DM signatures, 
see, for example, \cite{Bergstrom:1999jc,Donato:2003xg,Donato:2008jk,Fornengo:2013xda,Hooper:2014ysa,Bringmann:2014lpa,Giesen:2015ufa,Evoli:2015vaa}. 
DM constraints from CRs are, however, affected by uncertainties in the description of CR propagation in the Galaxy. 
So far,  DM limits have been derived for benchmark
propagation models, like the MIN/MED/MAX scenarios \cite{Donato:2003xg} obtained from observations of the Boron over Carbon (B/C) ratio. Such benchmark models introduce an order-of-magnitude uncertainty in the DM interpretation of CR fluxes.

The antiproton ($\bar{p}$) CR spectrum has recently been measured by the AMS-02 experiment with high precision \cite{Aguilar:2016kjl}.
In \cite{Cuoco:2016eej,Cuoco:2017iax,Cuoco:2019kuu} the AMS-02 measurements has been
used to reevaluate  the DM constraints in the light of the new data.
Thanks to the new precise data, antiprotons can be used not only to search for DM signatures but also to update the propagation model
itself~\cite{Korsmeier:2016kha}, without the use of B/C data. This approach has been indeed pursued in \cite{Cuoco:2016eej,Cuoco:2017iax,Cuoco:2019kuu}
where a joint fit of DM and CR propagation has been performed.
Propagation is described via the standard diffusion equation  and it's solved using \textsc{Galprop} \cite{Strong:1998fr,Strong:2015zva}.
Details of the propagation model and the numerical analysis are discussed in~\cite{Korsmeier:2016kha,Cuoco:2016eej,Cuoco:2019kuu}.

An important systematic in the analysis of CR $\bar{p}$ is   the nuclear
$\bar{p}$ production  cross-section required to model the astrophysical $\bar{p}$, i.e.,  the background for DM searches, produced as secondaries from CR nuclei collision with gas in the MW disc. This cross-section can be measured in the laboratory
but it's still affected by sizable errors. 
Nonetheless, \cite{Cuoco:2019kuu}  has investigate this issue in detail, 
and finds  that the cross-section uncertainty has, after all, only a minor impact on the analysis results.
This is somehow in contrast with the results of \cite{Reinert:2017aga}, who claims that cross-section uncertainties
have a large impact. In the light of the analysis of \cite{Cuoco:2019kuu}, however, is more likely
that the different results found in \cite{Reinert:2017aga} are not due to the cross-sections but to different details of the analysis,
perhaps to the different treatments of CR propagation.

\begin{figure}[t]
\vspace{-1cm}
\centering
\includegraphics[width=0.49\textwidth]{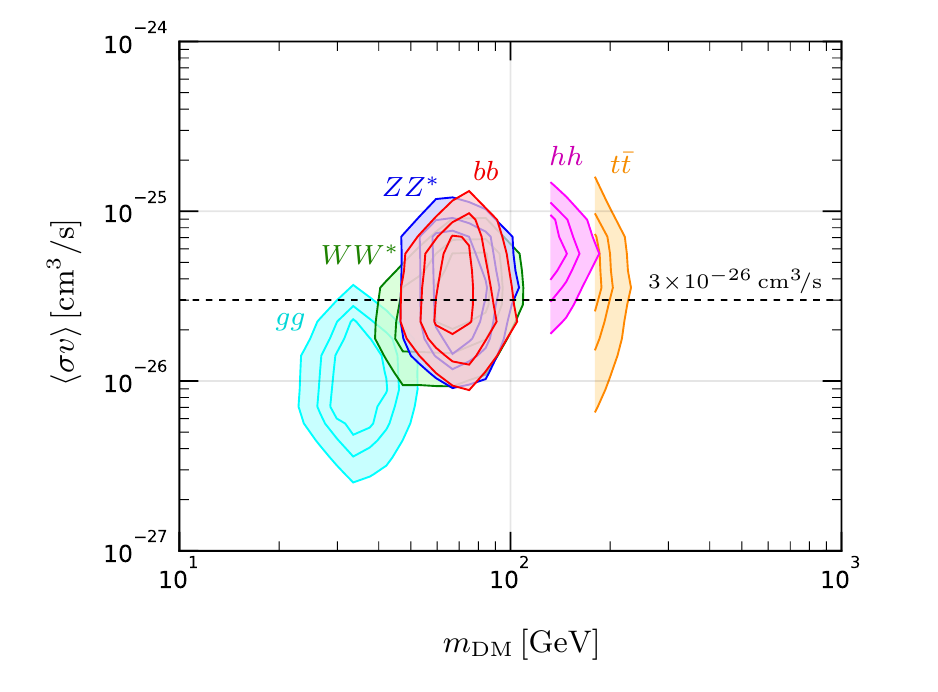}
\includegraphics[width=0.49\textwidth]{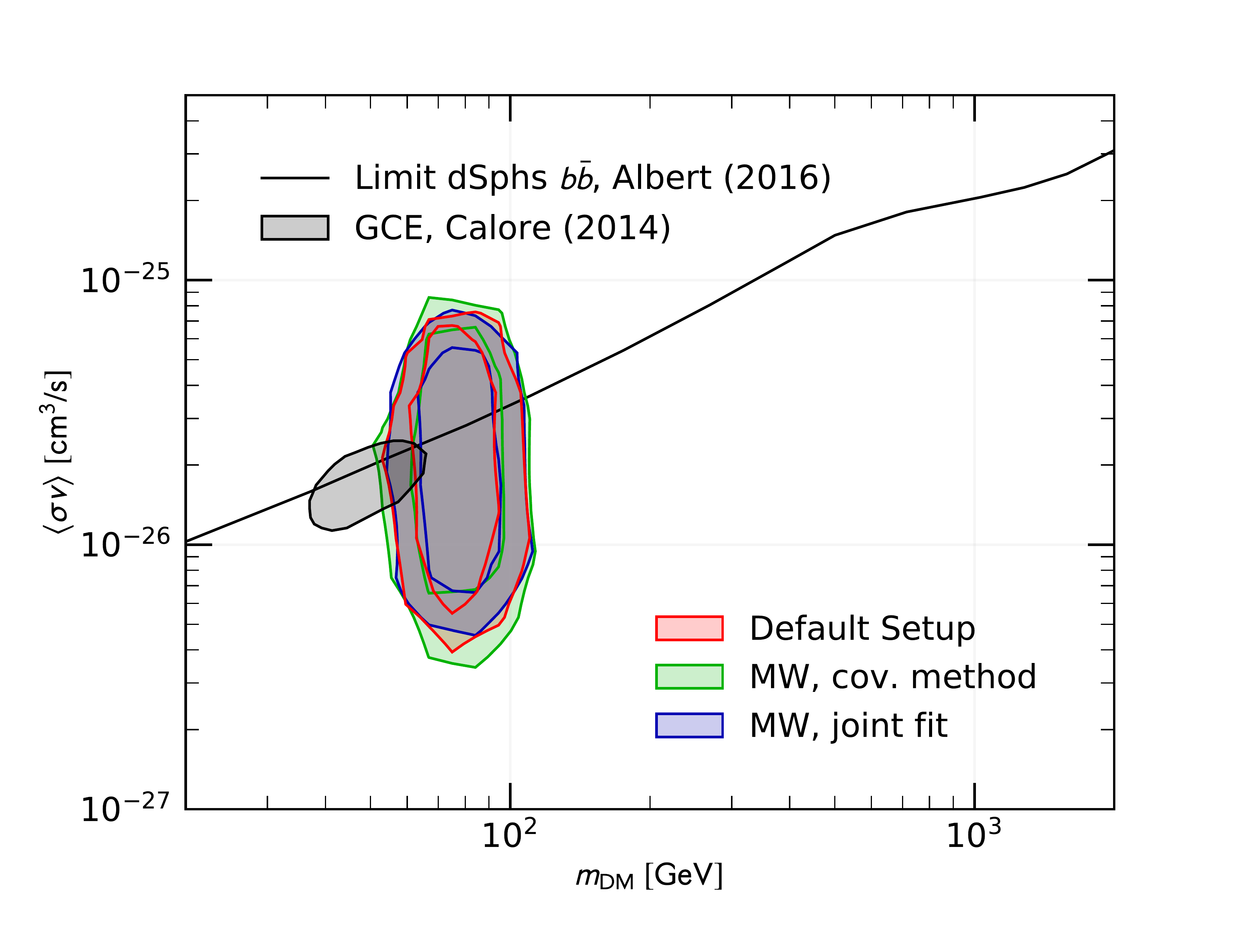}
\caption{{\it Left}: Cosmic-ray fit for the individual annihilation channels
in the $m_\text{DM}$-$\left\langle \sigma v \right\rangle$ plane, with 
1, 2, and 3\,$\sigma$ contours. The thermal cross section (dashed
horizontal line) is also shown. From \cite{Cuoco:2017rxb}. {\it Right}: effects on the DM contours for the $b\bar b$ channel
for different treatments of the nuclear cross-section uncertainties. Also shown are the constraints from dwarfs observations \cite{Fermi-LAT:2016uux},
and the region preferred by the fit of the GCE from \cite{Calore:2014xka}. See \cite{Cuoco:2019kuu} for more details.
}
\label{fig:CRind}
\end{figure}

Intriguingly, the analysis of \cite{Cuoco:2016eej,Cuoco:2017iax,Cuoco:2019kuu} finds that 
adding a DM component significantly improves the fit of the CR $\bar{p}$ data.
The DM mass-$\left\langle \sigma v \right\rangle$  region preferred by the fit for various annihilation channels is shown in Fig.~\ref{fig:CRind}.
The typical improvement in $\chi^2$ for the fit with and without DM is $\sim 12$, which corresponds to 
a significance of $\sim 3\, \sigma$.
The preferred DM mass is in the range \mbox{$\sim$ 30 GeV - 200 GeV} depending on the annihilation channel, while
for all the channels, the fit points to a thermal annihilation cross section $\left\langle \sigma v \right\rangle \approx 3 \times 10^{-26}$~cm$^3$/s. 
Complementary analyses  using propagation models fitted to the B/C data, have been performed in \cite{Cui:2016ppb,Cholis:2019ejx},
finding, similarly, an excess in antiproton  compatible with the above DM interpretation.

The right panel of Fig.~\ref{fig:CRind} shows the effects on the contours of different treatments of the nuclear cross-section uncertainties,
indicating, as mentioned above, that the effect is sub-dominant.
The figure also includes the region preferred by the fit of GCE from \cite{Calore:2014xka}, showing, interestingly
that the two DM hints from antiprotons and gamma rays are compatible.
A systematic investigation of the compatibility of the two hints in a formal joint CR+GCE fit 
with refitting of the GCE taking into account the uncertainty in MW DM profile has been performed in \cite{Cuoco:2017rxb},
and confirms  that there is a very good agreement between the two hints.


\subsection{Anti-Deuterium}
Beside nuclear cross-section uncertainties, other systematics which can affect the
interpretation of the $\bar{p}$ excess are modeling of the Solar modulation (i.e., CR propagation in the Solar system),
and the propagation model itself, which, although already sophisticate, might possibly still miss important ingredients.
A minor role should be played by possible correlations of the systematic errors in the experimental data points,
although a correct modeling could lead to more stringent constraints~\cite{Cuoco:2019kuu}.
A detailed study of these issues will be necessary to asses the robustness of the signal.
Alternatively, the DM interpretation can be investigated with  low energy ($\lesssim 1$ GeV/nucleon)
antideuterons $\bar{D}$~\cite{Aramaki:2015pii}. Astrophysical secondary $\bar{D}$ are expected to be produced similarly to 
secondary $\bar{p}$. The production threshold of $\bar{D}$ in CR collisions is, however, quite high, so that
production of low energy $\bar{D}$ is suppressed. On the contrary,  low energy $\bar{D}$ are expected from the annihilation/decay
of the, non relativistic, DM particle.  For this reason, the detection of low energy $\bar{D}$ would basically provide
a smoking gun in favor of the DM interpretation.
In particular,  AMS-02 and GAPS~\cite{Aramaki:2015laa} will be able to provide a crucial test of the $\bar{p}$ excess~\cite{Korsmeier:2017xzj}.

\section{Neutrinos}
Neutrinos, similarly to gamma rays, travel straightly from the sources and are, thus, a clean probe of DM.
The same targets analyzed in gamma rays (GC, dwarfs, etc.) can be investigated in neutrinos and give constraints on DM.
Differently from gamma rays, however, neutrinos are hard to detect and this translates into low
sensitivities, so that neutrino constraints on  $\left\langle \sigma v \right\rangle$ are typically various orders of
magnitude above the benchmark thermal reference value. 
The analogous of the gamma-ray EGB target  is DM searches in the neutrino diffuse emission. This is
a very interesting topic at the moment, since diffuse extra-terrestrial emission in the energy range $\sim$ 100 TeV -1 PeV has been recently detected
by IceCube for the first time~\cite{Aartsen:2013jdh}, and it has been suggested that part of this emission could be due
to decaying DM with masses also in  the 100 TeV -1 PeV range~\cite{Esmaili:2013gha,Feldstein:2013kka}.
This interpretation is not obviously ruled out although is subject to gamma-ray constraints~\cite{Cohen:2016uyg,Chianese:2019kyl}.
With IceCube collecting more statistics, the spectrum and  spatial distribution of the PeV diffuse emission
will become more clear along with its origin and, in turn,  this will clarify if a DM component is present.

Finally, a major target for DM searches in neutrinos is given by the Sun. In this case, annihilation
of DM captured and trapped in the core of the Sun would produce neutrinos detectable by IceCube. On the contrary, gamma rays and
CRs would remain trapped in the Sun itself, unless some further mechanism is invoked~\cite{Schuster:2009au}. 
There is, however, a background for this search, given by neutrinos produced in CR interaction with
the atmosphere of the Sun. IceCube is expected to become sensitive to this background soon~\cite{Ng:2017aur}. 
Once detected, further improvements in neutrino DM searches from the Sun will be limited.

\bibliographystyle{iopart-num}
\bibliography{CuocoTAUPbib}

\end{document}